\documentclass[journal=jacsat,manuscript=article,layout=twocolumn]{achemso}
\usepackage[version=3]{mhchem} % Formula subscripts using \ce{}
\usepackage{xr}
\usepackage[utf8]{inputenc}
\usepackage{graphicx}% Include figure files
\usepackage{hyperref}
\usepackage{caption}
\usepackage[normalem]{ulem}
\usepackage{verbatim}
\usepackage{url}
\usepackage[dvipsnames]{xcolor}
\newcommand{\ak}[1]{{\color{black}{#1}}}

\newcommand{\eb}{\textcolor{black}}
\newcommand{\ebII}{\textcolor{black}}
\usepackage{etoolbox}
\makeatletter
\patchcmd{\acs@contact@details}{E}{*\,E}{}{}
\makeatother

\title{Applied causality to infer protein dynamics and kinetics}

\author{Akashnathan Aranganathan}
\affiliation{Biophysics Program, University of Maryland, College Park, MD 20742, USA}
\altaffiliation{Contributed equally to this work}

\author{Eric R. Beyerle} 
\affiliation{Department of Biology, University of Copenhagen, 2200 Copenhagen N, DK}
\alsoaffiliation{Corresponding author}
\altaffiliation{Contributed equally to this work}
\email{eric.beyerle@bio.ku.dk}

\date{\today}

\abbreviations{MD, molecular dynamics; ML, machine learning; DTW, dynamical time warping; ACF, autocorrelation function; LE4PD, Langevin equation for protein dynamics; MSA, multiple sequence alignment; WT, wild type; MUT, mutant}

\begin{document}

\begin{tocentry}

\includegraphics[width=\linewidth]{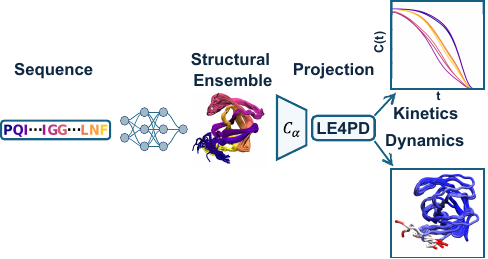}

\end{tocentry}

\begin{abstract}
The use of generative machine learning models, trained on the experimentally resolved structures deposited in the protein data bank, is an attractive approach to sampling conformational ensembles of proteins. However, the ensembles generated by these models lack timescale or causal information. We use the structural ensembles generated from AlphaFold2 at a range of MSA depths to parameterize the potential of mean force of an overdamped, memory-free, coarse-grained Langevin equation. This approach couples the AlphaFold2 ensembles to a causal model, allowing us to estimate the timescales spanned by the ensembles generated at each MSA depth. Performing this analysis on six variants of HIV-1 protease, we confirm an inverse relationship between MSA depth and the timescale of an ensemble's conformational fluctuations. The MSA depth essentially serves as a conformational restraint, and AlphaFold2 is generally able to probe timescales at or below those seen in microsecond-long, unbiased molecular dynamics simulations. We conclude by generalizing this approach to other generative structural ensemble-prediction methods \ebII{as well as co-folding models, in this case the biologically functional HIV-1 protease dimer}.
\end{abstract}

\section{Introduction}

It is well established that a protein's function cannot be elucidated directly from sequence alone, but rather requires a detailed knowledge of the conformational diversity encoded by the protein's free-energy landscape.\cite{Frauenfelder1991,Boehr2009,Ma1999,Henzler-Wildman2007,Copperman2015,Bouvignies2011} But, sampling effectively these conformations remains difficult, whether due to the limited temporal and spatial resolution offered by many biophysical experiments or issues regarding integration timescales and finite-size effects from computational simulations.\cite{Bottaro2018} While machine learning approaches have assisted both experiment\cite{Klukowski2025} and computation\cite{Noe2019} in populating the conformational ensemble of proteins, these approaches tend to lack information regarding the mechanisms and timescales of transitions between conformations.\cite{Kolloff2023} 

Typically, kinetic information at the residue level is gathered via long coarse-grained or atomistic molecular dynamics (MD) simulations.\cite{Lindorff-Larsen2011,Tozzini2007,Souza2025} Occasionally, the observed kinetics are augmented with time-resolved experimental information.\cite{Orioloi2020,Kuemmerer2021} However, although experimental techniques such as the T1 and T2 relaxation times from nuclear magnetic resonance (NMR) experiments inform on sub-microsecond motions,\cite{Clore1990a,Clore1990b,Schanda2025} they can lack atomistic and mechanistic information. 
Thus, we are beholden to trajectories generated from MD to access timescales at and below the microsecond, which encode the functional motions of proteins,\cite{Copperman2015,Henzler-Wildman2007} but obtaining sufficiently long simulations may require specialized resources.\cite{Shaw2021}

Biased MD can circumnavigate these walltime issues. This approach is system-dependent, requires prior knowledge, and alters the physical conditions to increase the sampling\cite{Henin2022}. Since the development of sequence-to-structure machine learning models such as AlphaFold2(AF2)\cite{Jumper2021} and RoseTTAFold\cite{Minkyung2021}, contemporary researchers utilize these models to generate conformationally diverse ensembles of proteins either by adding stochasticity or fine-tuning with MD data.\cite{delAlamo2022,Stein2022,Wayment-Steele2024,Zheng2024,Lewis2025} 
\ebII{However, to sample and describe accurately the dynamics of protein ensembles pulled from these machine-learned conformational ensemble generators, they must be rich in prior information regarding the physical laws and chemical interactions governing the functional motions of the protein and its surroundings.\cite{Zheng2023,Cui2025} }

Such ML models lack the causality mechanism provided by an MD integrator, meaning there is no inherent timescale attached to the generated ensemble. Therefore, any attempt to apply timescales to conformations generated from deep ML require coupling to either a deterministic (Hamiltonian) or stochastic (Langevin) integrator. Our primary goal here is to use the conformational ensembles obtained from these ML models to generate a harmonic potential of mean force matrix(PMF) to parameterize a memory-free, overdamped Langevin equation. Diagonalization of the PMF matrix yields an uncoupled set of Langevin modes, each of which are characterized by a specific time- and lengthscale.\cite{Caballero2007,Copperman2015,Beyerle2021,Beyerle2019} These Langevin modes yield autocorrelation functions (ACFs) for each residue in a protein.

Our approach is inspired by the framework proposed in Ref. \cite{Copperman2015}, where the authors map between a memory-free Langevin equation and a Gaussian network model using a potential of mean-force extracted from NMR conformations. This approach is linked closely to the diffusion maps method propounded in the manifold learning literature,\cite{Coifman2005, Coifman2008,Goekdemir2025,Rydzewski2023, Nadler2005} as well as its data-driven offshoots with applications to biomolecular dynamics.\cite{Noe2015,Evans2022,Rohrdanz2011,Rydzewski2023b} Our approach is different because we 1) use conformational ensembles generated from ML methods and 2) utilize the residue fluctuation basis set from Ref. \cite{Beyerle2021}, which incorporates anisotropy into the dynamics and furthermore maps into a principal component analysis (PCA) in certain limits. We test our methodology by examining the dynamics and kinetics of a set HIV-1 protease monomers consisting of both wild-type (WT) and mutant (MUT) sequences.

\begin{figure*}[ht]
    \centering

    \includegraphics[width=0.9\linewidth]{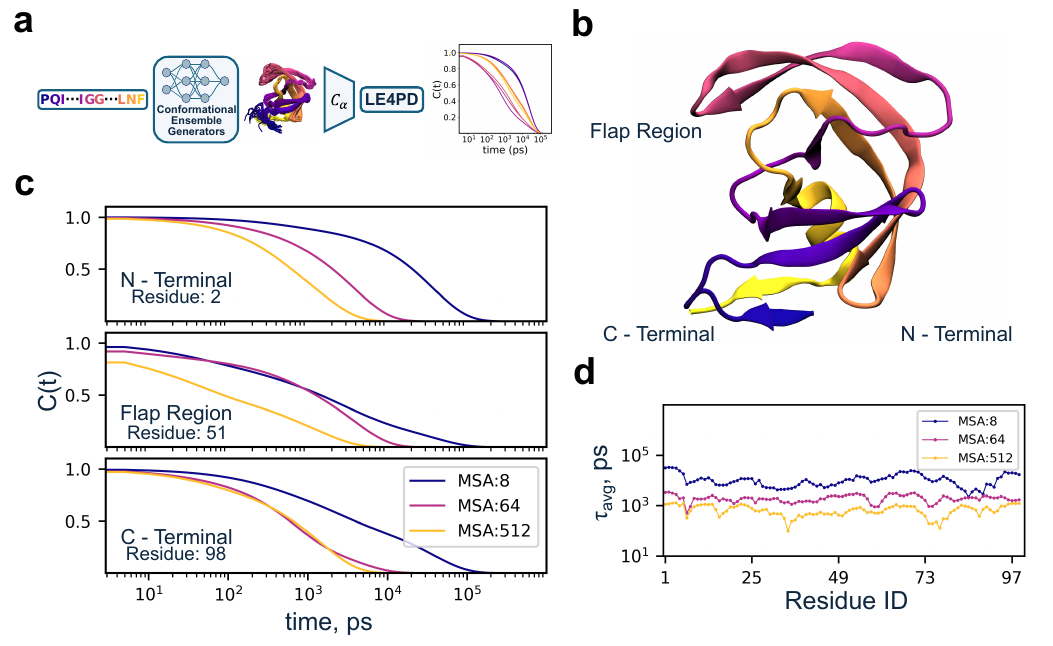}

    \hfill
    \caption{Overview of our method. a) The protein sequence is input to the machine learning architecture of choice, generating an ensemble of all-atom configurations. This ensemble is coarse-grained at the C$_{\alpha}$ level and input to the Langevin Equation for Protein Dynamics (LE4PD), which predicts the relaxation timescales (kinetics) of each residues' structural fluctuations (dynamics). b) A representative structure of the HIV-1 protease monomer is shown, with the flap (residues 49-51) and N- and C-termini labeled. c) Residue-dependent autocorrelation functions as a function of MSA depth input to AF2 for the three labeled structural regions in (b). d) Predicted relaxation time of each residue as a function of residue index for one HIV-1 protease sequence, PDB ID: 1EBW.}
    \label{fig:overview}
\end{figure*}

First, we use the reduced MSA (or MSA subsampling) approach to AF2 as the conformational generator to 1) demonstrate our pipeline and 2) find a relationship between reducing MSA information and the timescale associated with the generated ensemble. We generate one microsecond of atomistic MD simulation as a reference. By evaluating which rMSA is the closest, in the sense of our defined distance metric, to the reference MD, we can approximate the timescales spanned by an ML ensemble. The more robust result is mapping AF2 structures generated from an MSA depth of 8 between the relaxation timescales observed in the 100 ns and 1000 ns MD simulations. We find that AF2 ensembles generated using its default MSA cluster number of 512 (``full'' MSA depth) are comparable in timescale to those observed in a 10 ns simulation. We further demonstrate the generalizability of this protocol by comparing these results with other contemporary ML models \ebII{and applying the complete workflow to both a WT and MUT HIV-1 protease dimer using a co-folding model}.

\section{Results}

\subsection{Overview: A Map from Sequence to Kinetics}

An overview of our method of mapping from sequence space to a coupled dynamics-kinetics space provided by the solution to the ML-parameterized Langevin Equation for Protein Dynamics (LE4PD) equation is given in Figure \ref{fig:overview}a. Specifically, we apply this method on a set of six unique sequences encoding the HIV-1 protease monomer (structure shown in Figure \ref{fig:overview}b). The solution of the LE4PD gives autocorrelation functions (kinetics) for each residue and predicted root-mean-square fluctuations (RMSF) (dynamics). Depending on the hyperparameterization of the generative ML method, e.g. MSA depth to AF2, the kinetics of each generative ML ensemble is predicted using the LE4PD, as illustrated for three HIV-1 proteases residues (Figure \ref{fig:overview}c). We estimate the average decorrelation time $\tau_{\text{avg}}$ by integrating the predicted autocorrelation functions (ACFs) for each residue in the protein (Figure \ref{fig:overview}d).

\subsection{The Timescales of AlphaFold2-generated Ensembles are Inversely Correlated with MSA Depth}\label{sec:taus}

As the MSA depth in AF2 serves as a constraint on the predicted structures, as we decrease the MSA depth, we expect the ensemble's conformational variance to increase. Furthermore, since there is a general correlation between the variance of a protein's dynamics and the timescale of those dynamics,\cite{Beyerle2021} we expect that lifting the constraints via MSA depth reduction will increase the predicted residue relaxation timescales.

\begin{figure}[h!]
    \centering
        \includegraphics[width=0.48\textwidth]{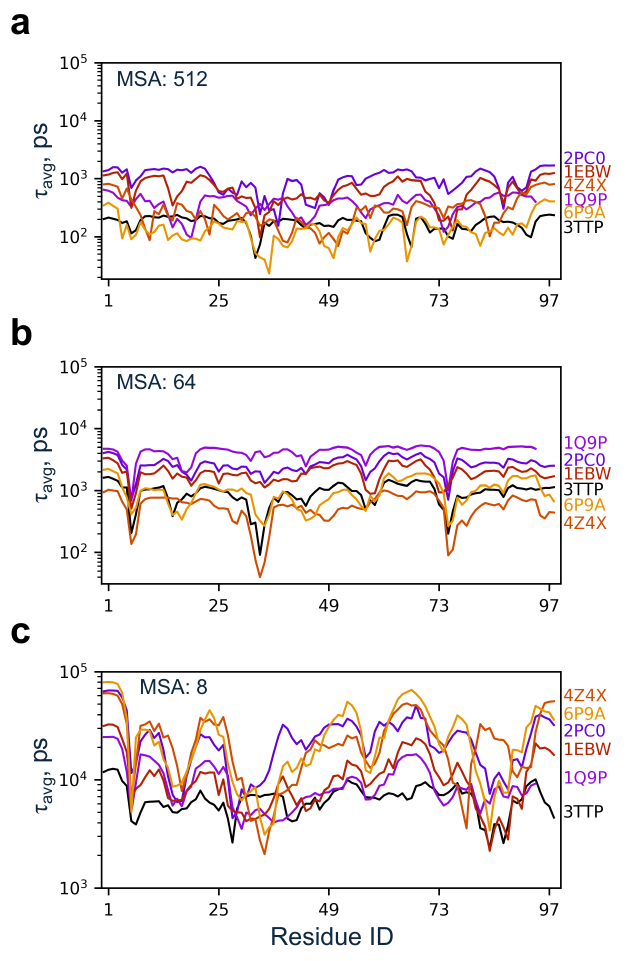}
        \caption{Estimated timescales for the relaxation kinetics of each residue for the six indicated sequences of HIV-1 protease at MSA depths of a) 512, b) 64, and c) 8. Each sequence variant label on the right-hand side of the plot is printed with the same color as its integrated timescale curve.}
    \label{fig:timescales}
\end{figure}

To measure this effect, in Figure \ref{fig:timescales} we plot the average decorrelation time, $\tau_{\text{avg}}$, for each residue, which is defined simply as the integral of the residue's ACF, indexed by $i$:

\begin{align}
\tau_{\text{avg}}(i)&=\int_0^{\infty} \langle\Delta \vec{R}_i(0)\cdot\Delta\vec{R}_i(t)\rangle dt.
    \label{eq:tau_avg2}
\end{align}
In Eq. \ref{eq:tau_avg2}, $\langle\Delta \vec{R}_i(0)\cdot\Delta\vec{R}_i(t)\rangle$ is the time correlation function for the $i^{th}$ $C_{\alpha}$ atom. The calculation of the time correlation function is in the Materials and Methods section.

Figure \ref{fig:timescales} plots the timescales for each sequence at three MSA depths: 512 (Figure \ref{fig:timescales}a), 64 (Figure \ref{fig:timescales}b), and 8 (Figure \ref{fig:timescales}c). When the MSA depth is at 512 and 64 (Figure \ref{fig:timescales}a and b), there is a clustering of either WT (1Q9P) or close-to-WT (five or fewer mutations from WT: 1EBW, 2PC0) at relaxation timescales slower than the highly mutated sequences (3TTP, 4Z4X, 6P9A), with 64 MSA depth showing a stronger separation. We observe a timescale separation between the near-WT cluster and the highly mutated cluster at certain stretches of the primary sequence, e.g., near the flap region (residues 49 through 51) at this highest MSA depth. 

However, at 8 MSA depth, there is a strong dependence of sequence composition on both the average relaxation timescale and the $\tau_{\text{avg}}$ of each residue individually. However, there is no correlation between a sequence's Hamming distance from the WT and the relaxation timescales when an MSA depth of 8 is used to parameterize the LE4PD. \ebII{Furthermore, there is no clustering seen in the equivalent relaxation timescale from 10, 100, and 1000 ns MD results, which are shown in the SI.} The SI \ebII{also} contains a more detailed analysis of per-residue timescales for all six systems. \ebII{From these results, the only definitive claim we can make is that, as the MSA depth is decreased, the timescales probed by AF2 show an increase, on average, in agreement with the equivalent MD results as the length of the MD simulation is increased. Unfortunately, due to the monomers' propensity to either dimerize spontaneously or form aggregates\cite{Ishima2001,Roesner2022}, there is no NMR relaxation data for comparing the mutant to wild-type relaxation timescales.}

%\sout{Could accelerated timescales in the mutants either play a role in their ability to evade pharmaceutical inhibition or show that AF2 has learned or memorized sequence patterns? We can speculate, but, since we do not investigate either ligand binding or interpretation of AF2 weights here, we can offer no definitive evidence either way. But,} 

\ebII{However,} since we see clustering at two higher MSA values studied (64 and 512) but not the lowest (8), we speculate this effect might be due to a tradeoff between the specific co-evolutionary signals or mutational information present in the MSA and the weights present in the AF2 architecture that allow it to fold proteins generally across a diverse sequence range. 

%Taking a free-energy perspective, we hypothesize the AF2 weights serve as an energy term while the MSA depth \ebII{is a reciprocal temperature}, with higher MSA pushing the predicted structures closer to the benchmark learned by the AF2 weights trained on PDB conformations. \ebII{Inspired by arguments from classical thermodynamics, we provide some evidence for this assertion in the SI by taking the average pLDDT score as a proxy for the free energy predicted by AF2 versus the logarithm MSA depth, which shows the concavity property.}

\subsection{MSA Depth is Inversely Correlated with the Span of the Slowest Collective Motions Generated by the AF2 Conformational Ensemble}

\begin{figure*}[htbp]
    \centering
    \includegraphics[width=0.9\linewidth]{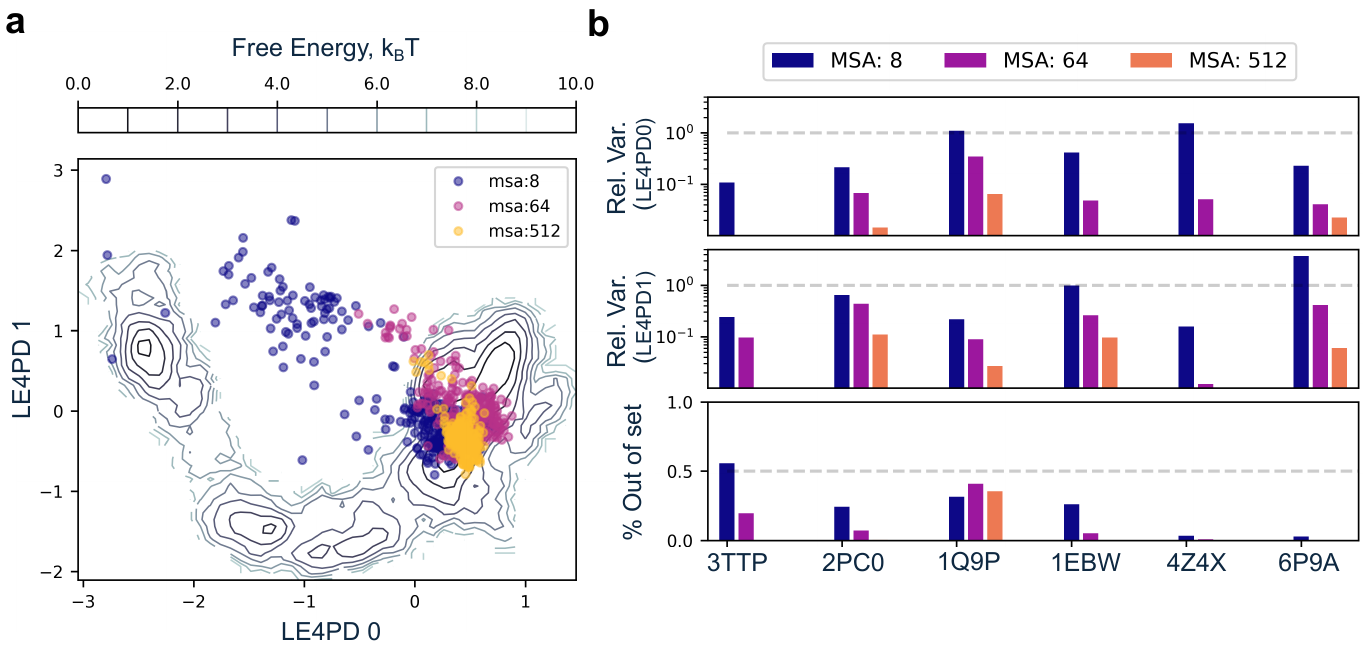}
    \hfill
    \caption{Measuring the volume of the space of the slow dynamics spanned by the AlphaFold2 conformational ensembles. a) AlphaFold2-generated ensemble projected onto the free-energy surface spanned by the two slowest LE4PD modes parameterized by a one microsecond MD simulation of the HIV-1 protease sequence encoded by PDB ID: 1EBW. b) Comparing the relative variance in the space spanned by the two slowest LE4PD modes of the AlphaFold2-generated ensembles compared to the one-microsecond MD simulation. We also show in the bottom plot what fraction of structures generated by AlphaFold2 at the indicated MSA depth are outside the support of the one-microsecond dynamics.}
    \label{fig:span}
\end{figure*}

The results discussed in the previous section show a general correlation of the relaxation time of collective motions and their variance. So, we expect that, as the MSA depth is decreased, and the amount of volume spanned by the AF2-generated ensemble increases, the timescales of the dynamics generally increase (Figure \ref{fig:timescales}) as well. To investigate this claim, Figure \ref{fig:span}a overlays the structures predicted at MSA depths of 8, 64, and 512 on the free-energy surface of the two slowest LE4PD modes calculated from a microsecond-long MD simulation starting from the structure encoded by PDB ID: 1EBW (simulation details are in the Materials and Methods section). Visually, as the MSA increases from 8 to 512, the predicted structures retreat into the global minimum on the free-energy surface, and the overall variance of the ensemble decreases (Figure \ref{fig:span}b). While the global minimum of both the AF2 and MD free-energy roughly coincide for 1EBW, this trend is not general (the SI shows equivalent scatter plots for the other sequences), and will almost certainly not hold for any given protein. Discrepancies between the modes of the ML- and MD-generated distributions are a function of the given sequence, ML architecture, and MD setup (force field, water model, etc.).

The bottom panel of Figure \ref{fig:span}b shows the fraction of out-of-set or outlying structures for each sequence at three MSA values. We define an outlying structure relative to the projection into the two slowest LE4PD modes calculated from the one microsecond simulation specific to each sequence. The outlying fraction of structures is a function of both sequence and MSA depth.

Generally, as the MSA is increased, the fraction of out-of-set structures decreases, which follows the argument that the MSA serves as a constraint keeping the predicted structures close to the training structures. The exception is 1Q9P, where the fraction of out-of-set structures is approximately constant as a function of MSA since, for this particular sequence, the predicted structures do not collapse to a minimum of the free-energy but rather near the node of the slowest LE4PD mode. This discrepancy could be due to the truncation in the C-terminus of 1Q9P. Since the truncated sequence is dominated by full-length sequences in AF2's training set, it might give the wrong mode for the predicted 1Q9P ensemble compared to MD, which simulates the truncated sequence explicitly.

Generally, the fraction of outlying structures should be a function of sequence, force field, simulation length, and subspace projection. For example, the sequences encoded in both 3TTP and 6P9A are more than 20 mutations from WT, but there are a far larger number of out-of-set structures in the generated ensemble using an MSA depth of 8 for 3TTP compared to 6P9A. In contrast, 2PC0 and 1EBW are both 5 mutations from WT, and they show a similar relative variance profiles and fraction of out-of-set structures. There does not appear to be a correlation between either the relative variance of the generated ensemble or the fraction of out-of-set structures as a function of Hamming distance from the WT sequence, which implies AF2 is not memorizing the dynamics of WT HIV-1 protease when inferring ensembles.

\subsection{Quantification of the Timescales Spanned by AF2 Evaluated at Different MSA Depths}

From Figure \ref{fig:span}a, it is clear that, at least for two sequences (e.g., 1Q9P and 4Z4X), the ensembles predicted from an MSA depth of 8 in AF2 structure generation yields an ensemble spanning the space of at least the slowest LE4PD mode, when measured using the fractional variance of that mode. However, assuming that the dynamics and kinetics of the entire protein will match just because the variance is approximately the same between projected AF2 and MD ensembles is questionable. So, we quantify how closely the residue-residue ACFs match between the AF2 and MD ensembles to determine rigorously the timescales of the ML ensembles' dynamics.

\begin{figure*}[ht]
    \centering
    \includegraphics[width=0.9\linewidth]{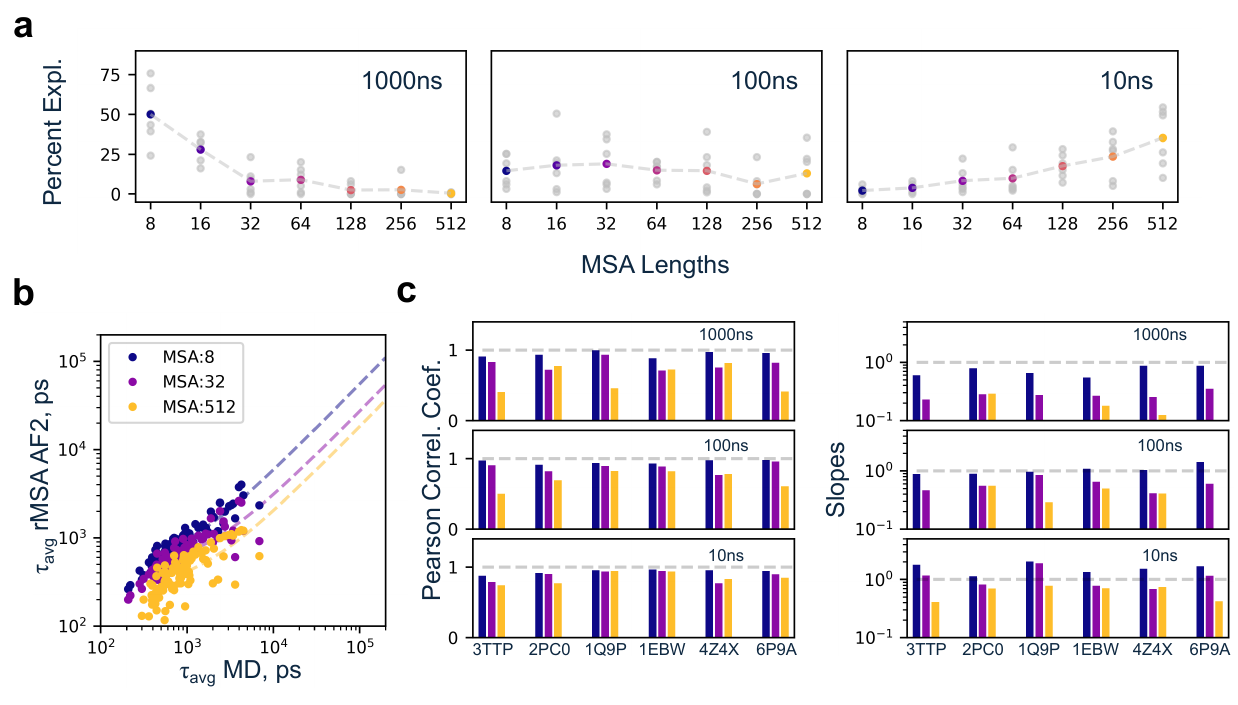}
    \hfill
    \caption{Reporting the timescales measured by AlphaFold2 conformational ensembles generated using different MSA depths. a) Percent of each residue's ACF best measured by each of the reported MSAs at three different timescales : 10, 100, and 1000 ns. Colored circles denote the average across sequences at each MSA while grey circle denote the result for each sequence individually. b) Scatter plot and linear fits (colored, dashed lines) of the average integrated correlation time from the reference MD simulations ($\tau_{\text{avg}}$ MD sim.) and the LE4PD theory parameterized using the AlphaFold2 conformational ensembles at the given MSA depth ($\tau_{\text{avg}}$ rMSA AF2). Reported are the correlation times from MD and AlphaFold2 for the HIV-1 protease sequence encoded in PDB ID: 1EBW at the 1000 ns timescale. c) Pearson correlation coefficient between the timescales reported in b), but for all six HIV-1 protease sequences studied here and for all three timescales (10, 100, and 1000 ns). Also reported in the second column are the slopes of the linear regression of $\tau_{\text{avg}}$ MD sim. onto $\tau_{\text{avg}}$ rMSA AF2 for all six sequences (1Q9P, 2PC0, 3TTP, 1EBW, 4Z4X, 6P9A) and all three timescales (10, 100, and 1000 ns). }
    \label{fig:AF2_timescales}
\end{figure*}

Figure \ref{fig:AF2_timescales}a plots the percentage of each protein's dynamics best explained by the AF2 ensemble generated at the indicated MSA depth. We map the ACFs using the DTW metric \eb{defined in the Methods section}. Nearly equivalent results are found using the KS instead; the KS results are displayed in the SI. Figure \ref{fig:AF2_timescales}a clearly shows the lowest MSA studied (8) is, on average, the best at describing the decorrelation timescales of the observed dynamics in the 1000 ns MD simulations. Conversely, using the full MSA depth of 512 allows for the best modeling of the decorrelation dynamics observed in the 10-ns-long MD simulations, on average. For the intermediate timescales in the 100 ns simulations, there is a broad range of well-performing MSAs from 16 through 128. Roughly, MSA depths between 8 and 16 model well the decorrelation timescales observed in 1000 ns MD simulations; MSA depths between 256 and 512 model best the decorrelation dynamics in 10 ns MD simulations. The in-between timescales in the 100-ns-long simulations are modeled well by MSA depths between 16 and 128.

As a second approach to quantify the predicted relaxation timescales between the coupled AF2-LE4PD approach and the benchmark MD simulations, Figure \ref{fig:AF2_timescales}b plots the mode-averaged residue decorrelation timescales defined in Eq. \ref{eq:tau_avg2} from the MD ACFs and the AF2-parameterized LE4PD's theoretical ACFs for the sequence in PDB ID: 1EBW. The scatter of points for each MSA is fit to a linear model. Figure \ref{fig:AF2_timescales}c reports the Pearson correlation coefficient and slopes of the linear models fit to each MSA for our six variants. 

Figure \ref{fig:AF2_timescales}c supports the results in Figure \ref{fig:AF2_timescales}a: the Pearson correlation coefficient for an MSA depth of 8 is closest to unity at the 1000 ns timescale, and it has the slope closest to unity in the linear fit at 1000 ns. \eb{Analyzing} the results for the 10-ns simulation timescale shows using an MSA depth of 512 gives a linear fit with slope closest to unity. In the limit that the Pearson correlation coefficient and slope approach unity, the predicted decay timescales \eb{(kinetics)} are identical, less an additive constant, which is why we use both measurables as goodness-of-fit metrics.

\subsection{Effect of Out-of-Set Sample Removal on the AF2-predicted Dynamics and Kinetics}

Generative ML models are prone to infer samples outside the support of known data distributions. For networks attempting to model physics, these output samples generally deviate from known physical reality by violating e.g. conservation of mass, energy, or both. For generative ML models of protein structures, such out-of-set samples\cite{Bubeck2023,Huang2025} may include, but are not limited to including, structures with unphysically large bond lengths, implausible torsion angles, or steric clashes.

Because \eb{we assign} such conformations null Boltzmann weight, it should be efficacious to remove these generated structures from downstream analyses. In theory, excising structures having infinite free-energy from the ensemble due to the above-mentioned modeling issues should yield a distribution closer to the target distribution, whether \eb{it's} from simulation or experiment. \eb{So}, pruning out-of-set structures should allow for a refine\eb{d ML ensemble} that aligns more closely with a target distribution. In practice, this amounts to a sharply defined maximum entropy approach\cite{Bottaro2018}.

\eb{Our target is} the free-energy surface spanned by the two slowest LE4PD modes parameterized using the one-microsecond MD simulations (Figure \ref{fig:span}). We remove the effect of the outlying structures by projecting the AF2-generated ensemble onto this free-energy surface, then remove from the generated ensemble any structures lying outside the space explored in the simulation. That is, we remove structures with infinitely high free energy i.e. zero Boltzmann weight. We parameterize the LE4PD using the modified AF2 ensemble and perform the calculations as done for the original ensemble. Finally, we examine how removing the out-of-set structures modifies the predicted RMSF profile, decorrelation times, and behavior of the ACFs.

\begin{figure*}[ht]
    \centering
    \includegraphics[width=0.8\linewidth]{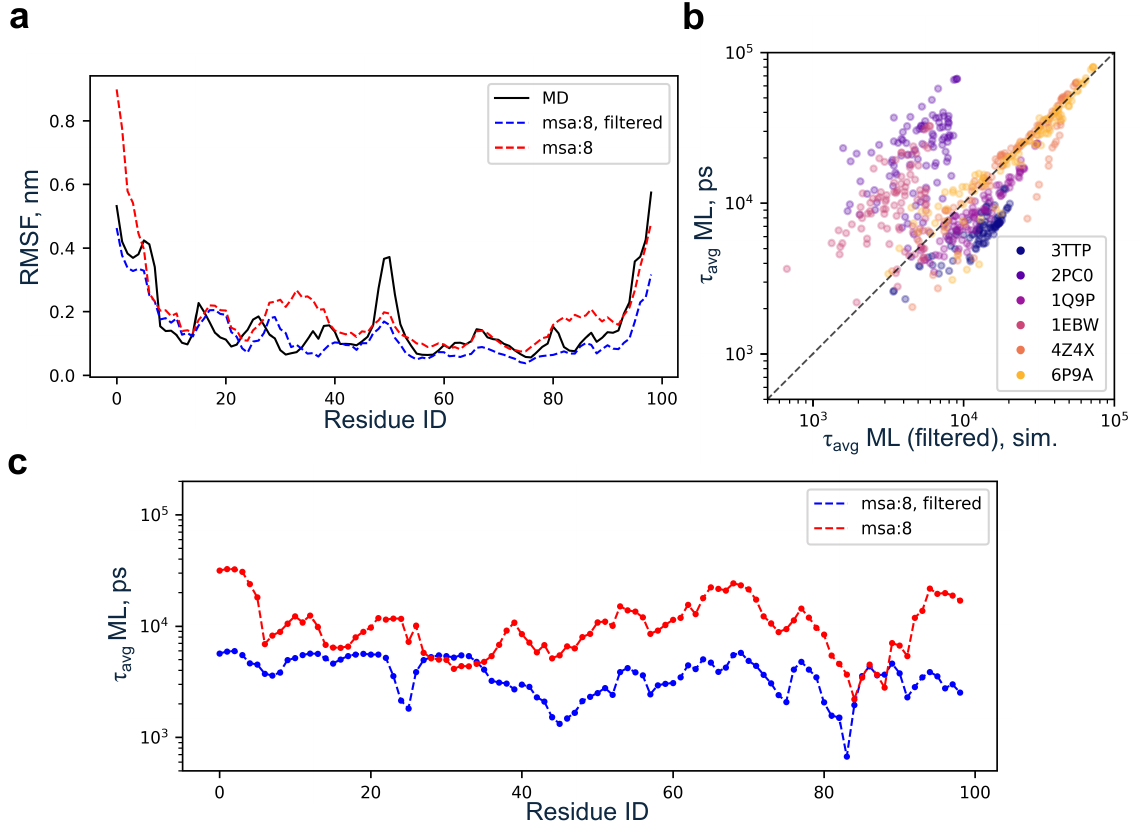}
    \hfill
    \caption{Effect of out-of-set structure removal on the predicted dynamics. All comparisons to MD are performed at the 1000 ns timescale. a) Root-mean-square fluctuations along the alpha-carbons predicted from the one microsecond MD simulation of HIV-1 protease with the sequence encoded by PDB ID: 1EBW (black) compared to the AlphaFold2 ensemble generated using an MSA depth of 8 and starting from the same sequence with (red, solid) and without (red, dashed) out-of-set structures in the ensemble input to the LE4PD theory. b) Correlation between the integrated correlation timescale predicted with the entire AlphaFold2 ensemble ($\tau_{\text{ML}}$) and the ensemble with out-of-set structures removed ($\tau_{\text{avg}}$ ML (filtered) for all six sequences. The points are colored by the sequence given in the subplot's legend. c) Integrated timescales for 1EBW $\tau_{\text{avg}}$ from the LE4PD theory parameterized using the AF2 ensemble generated using an MSA depth of 8 with (blue) and without (red) out-of-set structures included in the ensemble. Including the data points from all six sequences, the Pearson correlation coefficient between the timescales calculated from the ensembles with and without out-of-set structures is 0.912.}
    \label{fig:fraction_explained}
\end{figure*}

Figure \ref{fig:fraction_explained} shows the results after removing outliers. Figures \ref{fig:fraction_explained}a and \ref{fig:fraction_explained}b focus exclusively on the sequence encoded by 1EBW; equivalent plots for the other five sequences are given in SI Figures S9, S10, and S11. In Figure \ref{fig:fraction_explained}a, the RMSF from the AF2-generated ensemble with an MSA depth of 8 is compared to the reference one-microsecond MD simulation when out-of-set structures are either retained (dashed, blue line) or removed (dashed, red line). For this sequence, there is an almost universal decrease in the predicted RMSF across the entire primary sequence. In particular, there is a notable decrease in the RMSF near residues centered at index 30, with the removal of out-of-set structures pulling the RMSF in that region much closer to the microsecond MD. The same is true around the N-terminus. 

Figure \ref{fig:fraction_explained}b shows the correlation between the integrated timescales with out-of-set samples included (ordinate) or excluded (abscissa) for our six HIV-1 protease sequences, each demarcated by color. The sequences 2PC0 and 1EBW experience a significant decrease in timescale when out-of-set structures are removed while 4Z4X and 6P9A show little change in timescales; 3TTP and 1Q9P give, on average, slower timescales. Figure \ref{fig:fraction_explained}c shows the integrated timescale as a function of primary sequence for 1EBW, which shows clearly how its timescales decrease nearly uniformly as the outlying structures are removed.

Peering at Figures S9, S10, and S11 in the SI explains these disparate trends. Removing of out-of-set structures has a strong sequence dependence on the dynamics, i.e. RMSF, and kinetics in the form of $\tau_{\text{avg}}$. The six sequences can be divided into three classes: the first consists of 1EBW and 2PC0, where removal of outliers depresses the RMSF and timescales. The second is 4Z4X and 6P9A, where removal of outliers has a nearly negligible effect on the dynamics and kinetics. Finally, in the third is 1Q9P and 3TTP, where removing outliers actually increases the timescales nearly uniformly while the RMSF is not significantly altered. 

For this last class, the mean of the AF2-learned ensembles lies outside a mode of the simulation-parameterized LE4PD surface. Removing outliers eliminates structures near this mean, increasing the overall variance of the now-outlier-free distribution. Thus, where the predicted mean of the ML-learned ensemble does not overlap with an MD-predicted mode, there is a possibility that removing the outliers in the projected space will increase the variance, as well as, potentially, the RMSF and integrated timescales. When the ML-generated ensemble collapses to a mode of the projected MD-generated distribution, there will be both a reduction in RMSF and timescales because removing the outlying structures reduces the variance and hence the projected volume spanned by the ML-generated ensemble. When the distributions overlap, there will be few outliers, and their removal will have little effect on the results, as seen in the sequences encoded by 4Z4X and 6P9A. The specific sequence effects of outlier removal will depend on the reference free-energy surface.

\begin{figure*}[ht]
    \centering
    \includegraphics[width=0.9\linewidth]{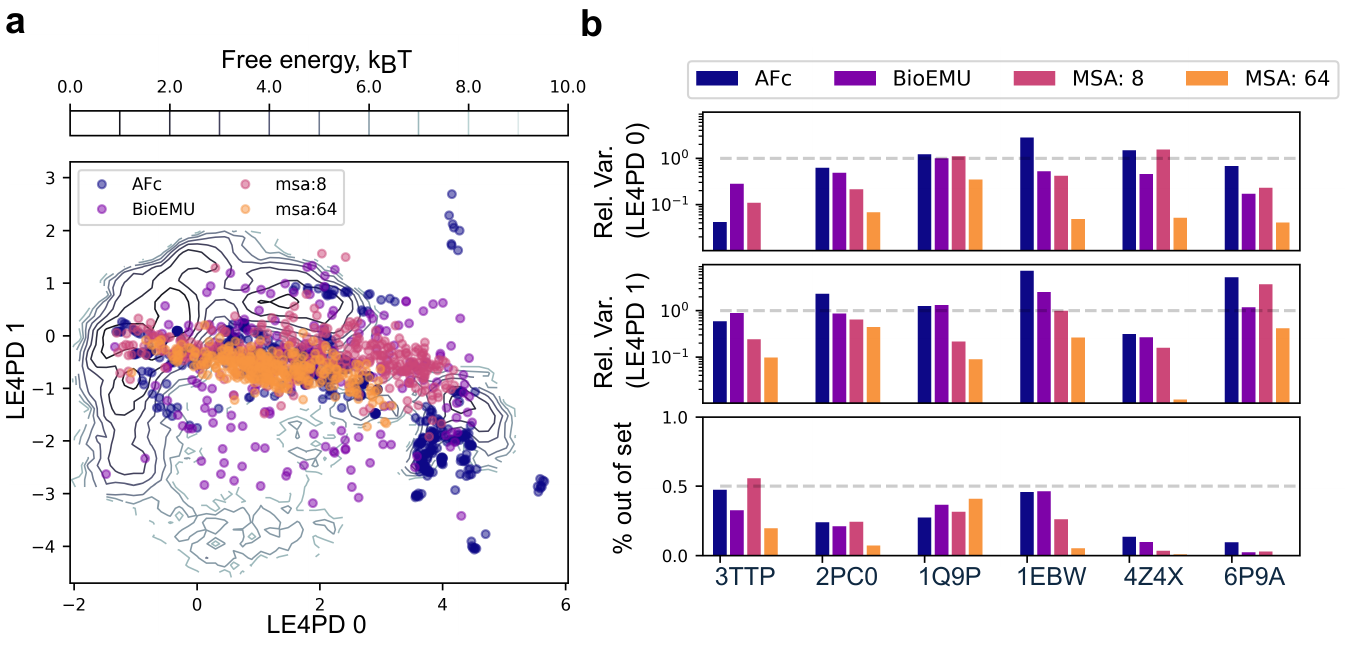}
    \hfill
    \caption{Measuring the volume of the space of the slow dynamics spanned by the AF2 conformational ensembles and other competing generative ML methods. a) Free-energy surface spanned by the two slowest LE4PD modes calculated from a one microsecond MD simulation of HIV-1 protease encoded by the sequence PDB ID: 1EBW. Projected onto the surface are the samples from AFc, BioEmu, and AF2 at different MSA depth. b) The relative variance of the two slowest LE4PD modes spanned by ensembles calculated using the same set of methods as in (a) for the six HIV-1 protease sequences. Also plotted in the bottom row is the fraction of out-of-set structures generated using each method. DiG results are omitted because they span a nearly negligible volume in this space, so the relative variance is not visible on the scale provided and none of the structures are outside the support of the one microsecond simulation.}
    \label{fig:genML_span}
\end{figure*}

\subsection{Comparison of AF2 Results to Other Generative Machine-learned Models}

Due to the plethora of methods currently available to generate conformational ensembles of proteins using ML techniques, we compare the AF2 results to three other contemporary state-of-the-art methods: the distributional graphformer (DiG),\cite{Zheng2024} AF-Cluster (AFc),\cite{Wayment-Steele2024} and the ``biomolecular emulator'' (BioEmu).\cite{Lewis2025}  

We choose the hyperparameters of the generative models to output 320 structures, the same size as the ensemble generated from AF2, although the exact size of the output ensemble depends on the ML architecture in use. For example, BioEmu automatically removes from the generated ensemble structures possessing steric clashes or bond length violations. For AFc, more than 320 structures may be output due to its clustering protocol. We solve the LE4PD equation separately for each sequence, using the ensemble generated by each method to predict dynamics and kinetics.

\begin{figure*}[ht]
    \centering
        \includegraphics[width=0.9\linewidth]{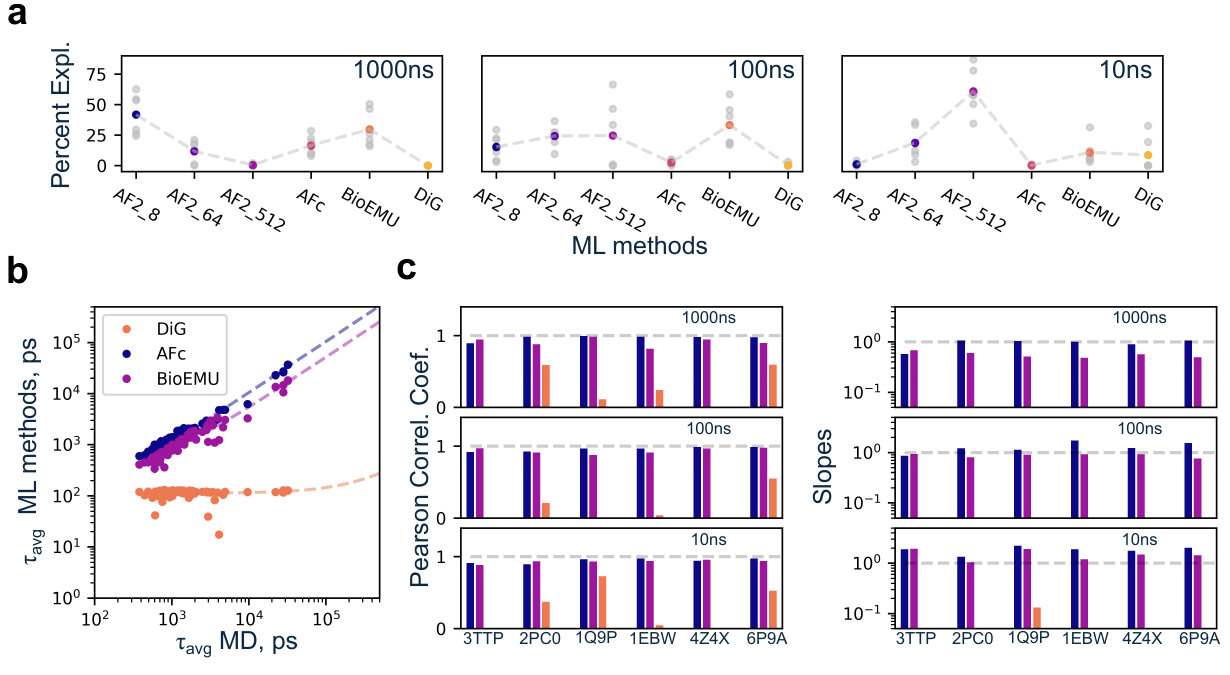}
        \caption{Reporting the timescales measured by alternative generative ML methods compared to AlphaFold2. a) Percent of residue ACFs best explained by generative ML methods for the HIV-1 protease sequence encoded by PDB ID: 1Q9P at the 10, 100, and 1000 ns timescales. Colored circles denote the average across sequences at each MSA while grey circle denote the result for each sequence individually. b) Correlation between the timescales($\tau_{\text{avg}}$) calculated from the ACFs from the one microsecond MD simulation and those calculated using the indicated ML ensembles input to the LE4PD theory. Dashed lines shown the best fit linear model to the data. c) Pearson correlation coefficients between the MD and ML integrated timescales for the three methods presented in (b) at the 10, 100, and 1000 ns timescales for the six HIV-1 protease sequences investigated here. Shown in the second column is the fit to the linear regression shown of $\tau_{\text{avg}}$ MD sim. onto $\tau_{\text{avg}}$ ML. The units for both $\tau_{\text{avg}}$ MD sim. and $\tau_{\text{avg}}$ ML are ps. }
    \label{fig:genML_timescales}
\end{figure*}

We can benchmark the predicted ML ensembles against the reference MD simulations. Figure \ref{fig:genML_span}a is analogous to Figure \ref{fig:span}a except the conformational ensembles from the ML methods in addition to AF2 are projected onto the free-energy surface spanned by the two slowest LE4PD modes. The span of each method is quantified, as before, by comparing the relative variance along the LE4PD modes individually in Figure \ref{fig:genML_span}b. We also display the fraction of structures that are located outside the span of the one microsecond MD ensemble in the bottom row of Figure \ref{fig:genML_span}b.

Figure \ref{fig:genML_span}a shows that the competing ML methods BioEmu and AFc span much larger volumes \eb{of the slow dynamical subspace} compared to AF2. This result is quantified in Figure \ref{fig:genML_span}b. However, as the last row of Figure \ref{fig:genML_span}b indicates, this large conformational volume can come at the cost of generating a large number of out-of-set structures. Strikingly, as shown in the SI, even though BioEmu can generate a relatively large number of structures that we quantify as out-of-set, removing these outlying structures has relatively little effect on either the dynamics or the $\tau_{\text{avg}}$ values. This result implies that BioEmu learns well the relevant structures in the folded ensemble of HIV-1 protease.

On the other hand, removing out-of-set structures from the AFc ensemble has a more noticeable effect on the predicted dynamics and kinetics. For all sequences, the predicted RMSF values remain greatly elevated nearly universally across the entire primary sequence for each variant even when outlying structures are removed. We conclude AFc is doing a poorer job compared to AF2 (at low MSA depths) and BioEmu at learning the folded free-energy landscape of the HIV-1 protease variants. The results for DiG are not shown because we find that the DiG structures span a negligible space around the crystal structure for all sequences, encoding essentially no functional dynamics.

Figure \ref{fig:genML_timescales} quantitatively evaluates the ML methods' performance. BioEmu and AFc describe the relaxation behavior of the protein's residues at the 1000 ns simulation timescale at a level that is similarly efficacious to AF2 with an MSA depth of 8 (Figure \ref{fig:genML_timescales}a). For modeling the dynamics in the 100 ns MD simulation, BioEmu is the most efficacious. At the 10 ns simulation timescale, using AF2 with an MSA depth of 512 is still the best model.

\begin{figure*}[htbp!]
    \centering
        \includegraphics[width=0.8\linewidth]{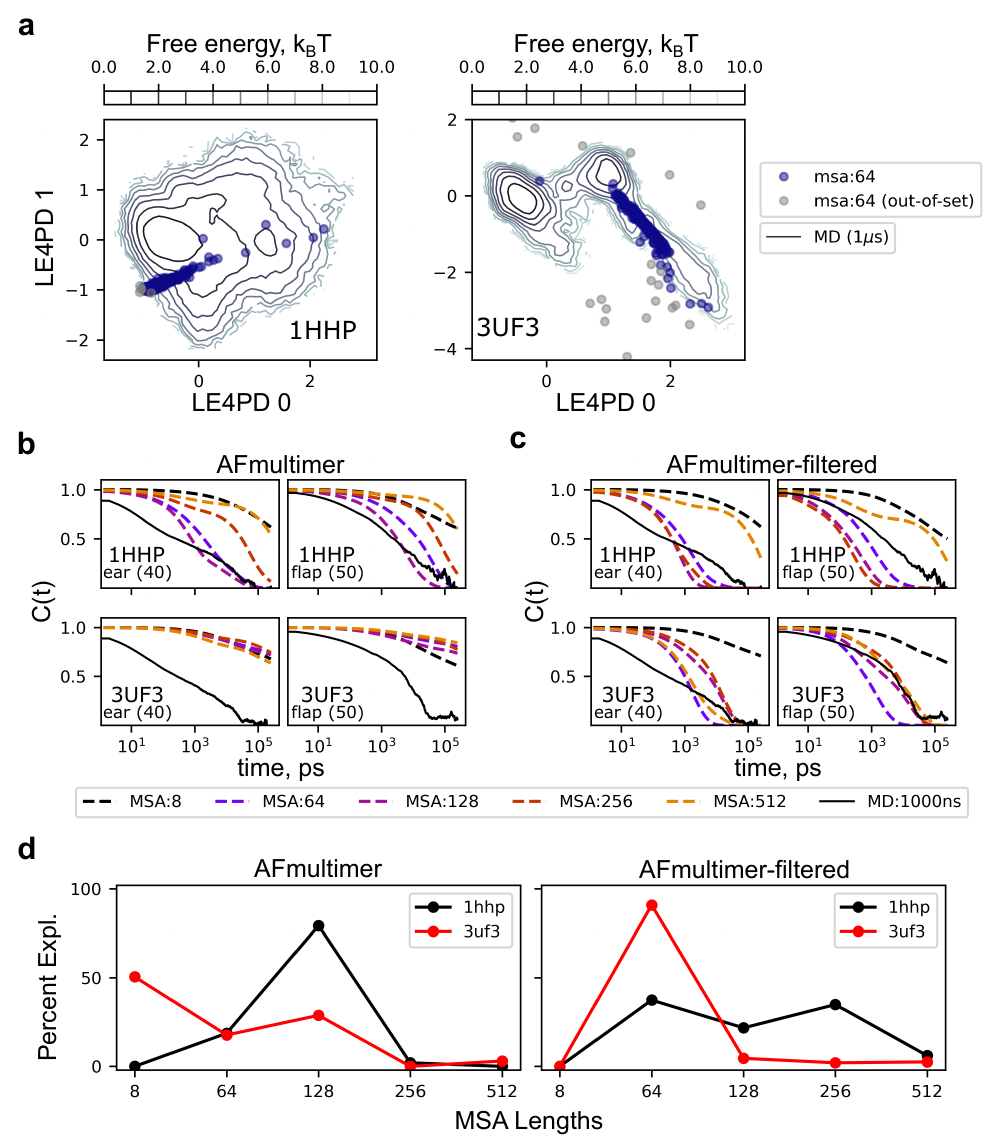}
        \caption{\ak{Applying our protocol for wildtype (1HHP) and a mutant (3UF3) HIV-1 protease dimer using AFmultimer. a) Projection of AFmultimer structures onto the free-energy surfaces spanned by the two slowest LE4PD modes predicted from a microsecond simulation started after equilibration of the structure deposited under PDB ID: 1HHP (left) or PDB ID: 3UF3 (right). b) Time correlation functions either calculated from MD (black curves) for the `ear' and flap residues in 1HHP (top panels) or 3UF3 (bottom panels). c) The same time correlation as in b), but with the predictions from the AFmultimer ensembles calculated using the reduced ensemble remaining after removing out-of-set structures. d) Comparison between ACFs from MD and our protocol for both pre- and post- removal of out-of-set structures from AFmultimer.}}
    \label{fig:multimer}
\end{figure*}

Figure \ref{fig:genML_timescales}b shows the correlation between the values of $\tau_{\text{avg}}$ of each residue in 1Q9P calculated from either the MD simulation or predicted from one of indicated ML methods. Again, we quantify the degree of correlation between the two sets of timescales using the Pearson correlation coefficient and the slope of the linear fit to the points in Figure \ref{fig:genML_timescales}c. The results quantify the statements made regarding the modeling efficacy of AFc and BioEmu: the Pearson correlation coefficient is near unity for all sequences at all MD simulation lengths for both methods. Furthermore, for the 1000 ns simulation, the linear fit to the AFc points has a slope near unity, again showing AFc integrated timescales are well-matched to the MD integrated timescales at this simulation length. While the BioEmu slopes differ from unity at the 1000 ns simulation length, they are much improved at modeling the dynamics in the 100 ns long simulations. For the 10 ns long simulations, the slopes for both AFc and BioEmu are significantly greater than unity, indicating an overprediction of the timescales.

We note the negative results when DiG models the dynamics from MD simulation. In Figure \ref{fig:genML_timescales}b, the DiG timescales are significant underestimates of the timescales seen in the 1000 ns MD simulation. This effect is quantified in Figure \ref{fig:genML_timescales}c, where the slope of the linear fits to the DiG integrated timescales is essentially invisible for all sequences and simulation lengths. These results indicate DiG is learning little to no protein dynamics and is instead returning ``ensembles'' \emph{very} tightly clustered around a single structure.

\subsection{\ak{Dynamics of HIV-1 dimer from AlphaFold-Multimer}}
\ak{HIV-1 protease functions only as a dimer, and the vast majority of the experimentally resolved structures are of the dimer. Due to spontaneous aggregation in solution, the only NMR ensemble of the monomer available in the PDB is for the 1Q9P sequence.\cite{Roesner2022,Ishima2001}} \ebII{Furthermore, computational studies have shown clear differences in the dynamics between the WT and MUT proteases, especially in the flap region, which are strong evidence for the differences in pharmaceutical efficacy between wildtype and mutant proteases.\cite{Chetty2016,Souffrant2023} As such, there have been many experimental\cite{Huang2014,Ishima1999,Kozisek2014,Heaslet2007,Shen2015,Kneller2020} and computational\cite{Perryman2004,Yu2017,Sankar2015} studies performed to study the function and dynamics of the dimerized HIV-1 protease.}

\ak{Thus, analysis of the dimeric HIV protease is important for predicting biologically relevant dynamics. In contrast to protein monomer ensemble prediction, ML modeling of multimer ensembles remains challenging.\cite{Lu2025} The nature of co-folding models can introduce artifacts and may result in spurious outputs. To avoid such complications, we avoid analyzing the dimeric HIV-1 protease using our proposed protocol as the core of this study. However, to provide a more complete picture of the predictions made by ML ensembles, we conclude this work by applying our workflow to wildtype (PDB ID: 1HHP\cite{Spinelli1991}) and a mutant (PDB ID: 3UF3\cite{Agniswamy2012}) HIV-1 protease dimer.}

\ak{We use the same MSA approach as for the monomers but use AlphaFold-Multimer (AFmultimer) \cite{Evans2021} instead of AF2 to generate an ensemble of 320 structures at each MSA depth. Since the reduced MSA approach reduces the number of sequence-based constraints, the co-folding models suffer more at lower MSA depths.\cite{heirarchical_af2rave} At lower MSA depths, the fold of the individual chains seems to be preserved; however, the binding orientation and the contact can be altered significantly. As these models are not causal and do not account for environmental conditions, the complex's structure prediction seems to be driven solely to place the two chains in close proximity. Overall, this reality suggests a better MSA depth to account for biologically relevant dynamics is higher than the monomeric case.}

\ak{To benchmark our results, we perform one-microsecond MD simulations starting from the corresponding crystal structures. A detailed analysis of our simulation is presented in the SI using a set of well-known collective variables\cite{Perryman2004} specific to this HIV-1 protease system. We also show that the slowest LE4PD modes capture the flap opening relevant to HIV-1 protease function. In Figure \ref{fig:multimer}, we demonstrate the results of our protocol applied to our selected dimers. Figure \ref{fig:multimer}a shows the overlap between the ensemble predicted by AFmultimer at an MSA depth of 64 and the MD distribution via projection onto the two slowest LE4PD modes from each simulation. Analysis using other MSA depths is shown in the SI. Even at higher MSA depths, AFmultimer appears to generate a significant proportion of structures that lie outside the MD-sampled region.}

\ak{Further structural analysis of ML ensembles revealed that AFmultimer prediction contained the dimer with C$_{2\text{h}}$ symmetry rather than the well-known experimentally resolved C$_2$ symmetry. The physical relevance of dimers with C$_{2\text{h}}$ symmetry is highly questionable, as it depends on the binding affinity and kinetics of that bound pose; however, C$_2$ symmetry corresponds to the protease’s functional assembly. Due to the presence of such structures in the ensemble, the ACF predictions are well exaggerated when compared to MD in Figure \ref{fig:multimer}b. For representative purposes, we chose flap tip residue 50 and `ear’\cite{Perryman2004} residue 40. A detailed analysis of other such residues is presented in SI.}

\ak{Following the reweighting method that penalizes out-of-set samples, we recalculated the ACFs for the reweighted ensemble. Figure \ref{fig:multimer}c clearly shows how the relaxation timescales move in line with those calculated from MD. We calculated the DTW metric defined in the Methods section between the predicted ACFs from our protocol, both the raw and reweighted (filtered) ensembles, and the ACF directly calculated from the MD simulation. Figure \ref{fig:multimer}d indicates that AFmultimer with an MSA length of 128-64 produces an ensemble that is closer to a 1 $\mu$s of MD simulation. Overall, these results: 1) support our initial hypothesis regarding the limitations of co-folding ML, like prediction of C$_{2\text{h}}$ symmetry dimers with high confidence; 2) highlight the requirement of a reduced MSA-type treatment to encourage the model to generate more possibly biologically relevant structures; and 3) demonstrate a requirement of a reweighting procedure. These results further show the differences in the MSA depth - timescales relation for monomer and multimer ML methods.}

\section{Discussion and Conclusions}

We provide a method qualitatively and quantitatively measuring the timescale of conformational ensembles output from a selection of deep, generative machine learning (ML) models for six sequence variants of human HIV-1 protease. We compute the timescales for the residue displacement autocorrelation correlation function (ACF) by assuming the backbone dynamics of the protein are governed by a coarse-grained, overdamped, and memory-free Langevin equation\cite{Beyerle2021} with a potential of mean force that is acceptably parameterized by the covariance matrix calculated from the generated ensemble. Diagonalizing the Langevin equation produces modes that form a basis set for the calculation of ACFs. This interface between generative ML techniques and the LE4PD methods allows for one-shot generation of dynamics and kinetics from sequence.

The residue-dependent timescales of the variants are benchmarked against each other to compare qualitatively the timescales of the ACFs as a function of sequence. This comparison can help determine the effect of mutations on the relaxation kinetics while the root-mean-square fluctuations calculated from the theory allow for the sequence-dependence of the dynamics. Quantitative benchmarking is performed by brute-force calculation of the analogous ACFs using molecular dynamics (MD) simulation trajectories of a given length. This second step of benchmarking against a known ensemble constructed from a simulation or experiment is not required for the predictive aspect of our method.

We show a correlation between reducing the MSA depth in AF2 for structure prediction and an increase in the timescales of the predicted ensemble due, at least in part, to an increase in the predicted variance of the ensemble along the slow dynamical modes. Reductions in timescale correspond to a reduction in the slow conformational dynamics as the depth of the MSA is increased (Figure \ref{fig:span}). Decreasing the MSA depth can increase the number of out-of-set structures in the projected LE4PD space (Figure \ref{fig:span}b). Removing outlying structures affects both the static properties (Figure \ref{fig:fraction_explained}a) and decorrelation timescales (Figure \ref{fig:fraction_explained}b). \ebII{We show these results hold for the biologically functioning dimer (Figure \ref{fig:multimer}) as well.}

Finally, we move beyond AF2 and analyze the conformational ensembles of three more recently developed architectures: AF-Cluster,\cite{Wayment-Steele2024} the distributional graphformer (DiG),\cite{Zheng2024} and the biomolecular emulator (BioEmu).\cite{Lewis2025} We find AF-Cluster predicts the slowest dynamics\eb{;} this prediction is due to a large number of outlying structures. BioEmu produces a much lower number of out-of-set structures, which makes its predicted decorrelation timescales more trustworthy compared to AF-Cluster, and the predicted backbone dynamics, as measured using the alpha-carbon root-mean-square fluctuation, are in excellent qualitative agreement with the 1 $\mu$s MD trajectories (Figure S13). Finally, we find that DiG samples only structures tightly clustered around an observed crystal structure, predicting virtually instantaneous and \eb{nearly frozen} conformational dynamics.

We anticipate the combined use of MD, the structural ensembles generated from ML, and the LE4PD to refine the weights of each ML structure in the ensemble in a manner more precise than that proposed here, \eb{e.g.} a maximum entropy approach.\cite{Bottaro2018} A re-weighting approach moves the sampled structures closer to a causal Hamiltonian or Langevin model at a given dynamical timescale and lengthscale with sequence specificity. Furthermore, the LE4PD outputs can be interfaced with the ML architecture's loss function to tune predicted ensembles in a subsequent round of training so that the model infers an ensemble with the desired dynamics. We hope this type of approach helps a move toward a causal generator of protein ensembles avoid\eb{ing the computationally expensive querying of }an integrator.

\section{Materials and Methods}

\subsection{Machine-learned Protein Structural Ensemble Generation}
Here are the details of the methods used to generate structural ensembles from all the machine learning methods we utilize.

\textbf{AlphaFold2\cite{Jumper2021,Mirdita2022}:} AlphaFold2 structures are generated at the specified MSA depth using LocalColabFold (\url{https://github.com/YoshitakaMo/localcolabfold}). For each sequence, we use three recycles, 64 seeds, and dropout. We raster the MSA from 8:16 until 512:1024 (``full MSA'') on a base-2 logarithmic scale. The structures are ranked using the pLDDT score. The input script we use to generate AF2 structures is given in the GitHub repository affiliated with this manuscript.

\textbf{AF-Cluster\cite{Wayment-Steele2024}}: Structures are generated using the Colab notebook present in the GitHub repository (\url{https://github.com/HWaymentSteele/AF\_Cluster}) with the appropriate HIV-1 protease sequence inserted instead. We use the mmseqs2 method to generate the MSA with $\mathsf{pair\_mode}$ set to $\mathsf{unpaired\_paired}$. We set the number of copies equal to 1; we specify that the minimum number of samples in a cluster is 10; and we do not use any templates.

\textbf{DiG\cite{Zheng2024}}: We use the default parameters specified in the original manuscript and on the project's GitHub repository (\url{https://github.com/microsoft/Graphormer/blob/main/distributional\_graphormer/protein/run\_inference.py}), except that we change the input files to 1) a Python pickle file containing the full MSA and 2) the FASTA file containing the HIV-1 protease sequence of interest. As with AlphaFold2, we specify that DiG output 320 structures during the inference step.

\textbf{BioEmu\cite{Lewis2025}}: We run BioEmu with the default parameters specified in its GitHub repository (\url{https://github.com/microsoft/bioemu}), except we demand it generate 320 structures in the inferred ensemble. This number of structures is the upper bound for the number in the final ensemble, as BioEmu then proceeds to pass the structures through a filtering procedure removing physically unallowable configurations from the output.

The six HIV-1 protease sequences are taken from the following PDB identifiers: 3TTP\cite{Kozisek2014}, 2PC0\cite{Heaslet2007}, 1Q9P\cite{Ishima2003}, 1EBW\cite{Andersson2003}, 4Z4X\cite{Shen2015}, and 6P9A\cite{Kneller2020}. We choose to simulate the HIV-1 protease monomer because 1) it is one of the test systems in Ref. \cite{Copperman2015} and 2) it is prone to variant selection due to resistance mutations, which also have notable effects on protease's structure and dynamics.\cite{Piana2002, Galiano2009,Perryman2004,Appadurai2016,Souffrant2023,Liu2008} While functionally active only in the dimerized form, we believe that studying the monomer in isolation can still yields insights regarding the protease's functional behavior. 

\ebII{MD simulations of the dimers were run using the same protocol as the monomer simulations. The starting structure for the WT dimer is PDB ID: 1HHP\cite{Spinelli1991} and PDB ID: 3UF3\cite{Agniswamy2012} for the mutant. To generate AF2 equivlent structures of the dimers, we utilize AlphaFold-multimer (AFmultimer)\cite{Evans2021}. We use the ColabFold implementation (\url{https://github.com/sokrypton/ColabFold}).}

\subsection{Generating Ground-Truth Conformational Ensembles Using Atomistic Molecular Dynamics}

To generate a reference conformational ensemble encoding dynamics of each HIV-1 protease variant, we perform atomistic, explicit solvent, classical MD simulations using the OpenMM package\cite{Eastman2017} with the amber99sb-ildn\cite{Lindorff-Larsen2010} and the TIP3P water model.\cite{Jorgensen1983} Each simulation is started using the AF2 structure with the highest predicted confidence. The protein is solvated, ionized until neutrality, equilibrated for 300 ps in the NVT ensemble with position restraints, 300 ps in the NPT ensemble with position restraints, and 300 ps in the NPT ensemble without position restraints; all the equilibration steps are performed at a temperature of 300 K. Finally, a one microsecond production run in the NPT ensemble is performed at 300 K, which is then analyzed, including the calculation of ACFs. For the analysis at 10 and 100 ns, the first 10 and 100 ns of the simulation are extracted and analyzed, respectively. For all MD simulation runs, we use the Langevin integrator. The temperature is controlled using a Nose\'{e}-Hoover thermostat\cite{Nose1984,Hoover1985} in all cases and, for the NPT simulations, the pressure is controlled using the Parinello-Rahman barostat.\cite{Parrinello1981} Detailed scripts for reproducing the OpenMM simulations, including the energy minimization and equilibration steps, are given in the GitHub repository for this manuscript (\url{https://github.com/erb24/af2-dynamics}). 

\subsection{Using Conformational Ensembles to Parameterize a Langevin Equation to Approximate Conformational Dynamics}

The fundamental assumption in this paper is the residue-level fluctuation dynamics of folded, solvated proteins at the microsecond scale and faster obey a memory-free, overdamped Langevin equation\cite{Beyerle2021}:
\begin{equation}
\frac{d\Delta \vec{R}_i(t)}{dt} = -\frac{3k_BT}{\overline{\gamma}}\sum_{j,k}H_{i,j}A_{jk}\vec{R}_k(t) + \vec{\zeta}_i(t),
    \label{eq:le4pd}
\end{equation}
where $\vec{R}_i(t) = (x_i(t)-\langle x_i\rangle,y_i(t)-\langle y_i\rangle,z_i(t)-\langle z_i\rangle)^T$ is a vector measuring the fluctuations of the Cartesian coordinates of residue $i$ from its equilibrium position $\left(\langle x_i\rangle,\langle y_i\rangle,\langle z_i\rangle\right)^T$, $k_B$ is the Boltzmann constant, $T$ is the temperature in Kelvin, $\overline{\gamma}=1/N \sum_i \gamma_i$ is the average friction coefficient, and $\vec{\zeta}_i(t)$ is a white noise term modeling solvent collisions that obeys the following fluctuation-dissipation theorem: $\langle \vec{\zeta}_i(t) \cdot \vec{\zeta}_j(t^{\prime})\rangle = 6k_BT\delta_{i,j}\delta_{t,t^{\prime}}$. Finally, $H_{ij}$ is the $i,j^{\text{th}}$ element of the hydrodynamic interaction matrix giving long-ranged hydrodynamic interactions between residues $i$ and $j$ while $A_{ij}$ is the $i,j^{\text{th}}$ element of the potential of mean force matrix giving the pairwise interactions between residues $i$ and $j$, which are, in general, long-ranged. This potential of mean force matrix is defined as the pseudo-inverse of the covariance matrix, $\mathbf{C}:=\langle \left(\mathbf{R} - \langle \mathbf{R}\rangle\right)^T\left(\mathbf{R}-\langle \mathbf{R}\rangle\right)\rangle$ in the space spanned by the residue trajectories\cite{Beyerle2021}: $\mathbf{A}:=\mathbf{C}^{\dagger}$. For more detailed explanations of the $\mathbf{H}$ and $\mathbf{A}$ matrices, the reader is invited to read Ref. \cite{Beyerle2021} and its associated supplementary material.

The coupled dynamics at the residue level is separated by applying a similarity transformation to Eq. \ref{eq:le4pd}, giving the following equaiton of motion\cite{Beyerle2021}:
\begin{equation}
\frac{d\vec{\xi}_a(t)}{dt} = -\sigma_a\vec{\xi}_a(t) + \vec{\zeta}_a(t),
    \label{eq:le4pd_modes}
\end{equation}
with $\vec{\xi}_a(t)=\sum_i Q^{-1}_{ai}\Delta \vec{R}_i(t)$, $\lambda_a:=\left(\mathbf{Q}^{-1}\mathbf{H}\mathbf{A}\mathbf{Q}\right)_{aa}$, $\sigma_a:=\frac{3k_BT}{\overline{\gamma}}\lambda_a$, and $\vec{\zeta}_a(t) = \sum_{i}Q^{-1}_{ai}\vec{\zeta}_i(t)$ is the white noise term in the space of Langevin modes. The friction coefficients $\gamma_i$ for each residue are calculated using the solvent accessible surface area and assuming it decomposes into solvent-dependent and a solvent-independent terms.\cite{Beyerle2021}

 \subsection{Calculating Time Correlation Functions from Structural Ensembles}
 
The mode solutions to the Langevin equation given in Eq. \ref{eq:le4pd_modes} decay exponentially with a characteristic timescale given by $\tau_a=\sigma_a^{-1}$ and lengthscale $l_a=\mu_a^{-\frac{1}{2}}$, and the ACFs in the space of residue fluctuations is given as a linear combination of the ACFs for the Langevin modes:
\begin{equation}
\langle\Delta \vec{R}_i(\tau)\cdot\Delta\vec{R}_i(\tau+t)\rangle=\sum_{a > 6} \frac{Q_{ia}^2}{\mu_a} e^{-\frac{t}{\tau_a}}.
    \label{eq:tcf}
\end{equation}
In Eq. \ref{eq:tcf}, the summation skips the first six modes as they correspond to the three translational and three rotational modes of motion, which are irrelevant for modeling the internal conformational dynamics of interest here. \eb{Furthermore, $\mu_a:=\left(\mathbf{Q}^{-1}\mathbf{A}\mathbf{Q}\right)_{aa}$.} The integral of eq. \ref{eq:tcf}, which is eq. \ref{eq:tau_avg2}, can be written in the mode coordinates as follows:
\begin{align}
\tau_{\text{avg}}(i) &=\int_0^{\infty}\sum_a {Q_{ia}}^2\mu_a \exp\left[-t/\tau_a\right]dt/\sum_a {Q_{ia}}^2\mu_a \notag\\
&= \sum_a {Q_{ia}}^2\mu_a \tau_a / \sum_a {Q_{ia}}^2\mu_a =\langle \tau_a\rangle_i.
    \label{eq:tau_avg1}
\end{align}
Defining the weight of mode $a$'s contribution to $\tau_{\text{avg}}(i)$ as $w_a:={Q_{ia}}^2\mu_a$ is used in transitioning from the left to the right side of the second equality in  Eq. \ref{eq:tau_avg1}. As such, $\tau_{\text{avg}}(i)$ is the average mode decorrelation time for residue $i$, as predicted using the LE4PD. 

Finally, since the Langevin equation is coarse-grained, it displays accelerated dynamics on the correspondingly smoothed free-energy surface. \cite{Lyubimov2011} To account for this speed-up, we approximate the missing barriers \emph{a posteriori}\cite{Zwanzig1988} using an approximate scaling law for proteins derived in Ref. \cite{Copperman2017}: $\tau_a \rightarrow \tau_a \exp[\epsilon\sqrt{\lambda_a}/k_BT]$, where $\epsilon:=6.5$ kcal / (mol nm) is an energy per unit length. We use this approximation since it is found, in Ref. \cite{Copperman2017}, to accurately describe the timescales of protein dynamics between the sub-$\text{\AA}$ngstrom to the nanometer lengthscales, which are the same as those examined here.

\subsection{Measuring the Distance between Time Correlation Functions}

To quantify the distance between the MD- and AF2-generated ACFs, we use two distinct methods: the Kolmogorov-Smirnov (KS) test\cite{Siegel1956} and dynamical time warping (DTW).\cite{Ray2024,Meert2020} The details on how these methods are used are given below.

To perform distance calculations between ACFs using the KS test, each ACF is converted into its corresponding survival function, $S(t)$. Defining the ACF $C(t):=\langle \Delta \vec{R}_i(\tau) \cdot \Delta \vec{R}_i(\tau + t)\rangle$, we have $S(t):= 1 - C(t)$. This survival function is equivalent to the cumulative distribution functions normally input to a two-tail KS test, and it is used to calculate the usual KS distance between the two survival functions: $d_{\text{KS}}=\sup_t |S_{\text{MD}}(t) - S_{\text{AF2}}(t) |$.  

Similarly, we use DTW in the usual manner: we calculate a mapping between the MD- and AF2-generated ACFs such that the path-warping distance is minimized: $d_{\text{DTW}}(C_1,C_2)=\min_p(c_p(C_1,C_2))$, given that $p$ is a suitably defined warping path between ACFs $C_1$ and $C_2$ while the cost of an suitable warping path between $C_1$ and $C_2$ is given by $c_p(C_1,C_2)=\sum_{l=1}^{L(p)}c(C_{1}(l), C_2(l))$. In calculating the cost for taking an individual step along the warping path $c(C_1(l),C_2(l))$, we use only a single time series index $l$ since the compared ACFs from simulation and theory are of the same length and with the same lag between each sequential value.

The percentage of each protein's dynamics best explained by the AF2 ensemble at the indicated MSA depth is calculated using the DTW distance matrix $\mathbf{D}$, where $\mathbf{D}_{m,n}$ represents the DTW distance between the ACF calculated by brute-force from the MD and from the LE4PD theory parameterized using the AF2-generated ensemble at MSA depth $m$ for residue $n$ (these matrices are shown in SI). This matrix is calculated at the indicated timescale. The percent explained is calculated as:
\begin{align*}
    \mathbf{k} &= \arg\min_m \mathbf{D}_{m,n},
\end{align*}
 where $\mathbf{k}$ is a vector of length $N$ that contains the values of MSA depth at which the ACF from the LE4PD is closest to the MD, $N$ is the number of residues, and $m \in \left\{8, 16, 32, 64, 128, 256, 512\right\}$. Then
\begin{align*}
    \text{\% explained at $m$} &= \frac{100}{N}\sum_{n=1}^N \delta({\mathbf{k}_n -m)},
\end{align*}
which counts the percentage of residues' ACF best matched by LE4PD parameterized from AF2 ensemble at MSA depth $m$. This calcualtion was performed for all three MD timescales (10, 100, and 1000 ns). 

%TC:ignore

\section*{Acknowledgments}
We thank Prof. Pratyush Tiwary and the University of Maryland for providing the funding and access to the computational resources required to complete this project. Specifically, the MD simulations and generative modeling are performed using the Zaratan cluster at the University of Maryland, College Park; the bridges2 cluster at the Pittsburgh Supercomputing Center; and the Biowulf cluster at the NIH campus in Bethesda, MD. A.A. thanks NCI-UMD Partnership for Integrative Cancer Research for fellowship. We thank Prof. Marina Guenza (University of Oregon) for suggesting a comparison of the dynamics from the 10 ns through microsecond scales and Dr. Julian Streit for a critical reading of the manuscript.

\begin{suppinfo}

Detailed analysis of the backbone time correlation functions and further analysis of the ensemble sampled for all the systems can be found in the supplement.

\section{Data and Software Availability}
All data associated with this work is available through \url{https://github.com/erb24/af2-dynamics}. The code used for the LE4PD approach of AF2 and other ML-method generated ensembles is available at \url{https://github.com/erb24/af2-dynamics}. Codes and parameters used to specifically run the simulations can be found at \url{https://github.com/erb24/af2-dynamics}.

\end{suppinfo}

\bibliography{jcim/ref_jcim_noURL}
% \quickwordcount{jcim/main_jcim}
% \detailtexcount{jcim/main_jcim}
%TC:endignore

\end{document}